\documentclass[iop]{emulateapj}
\pdfoutput=1
\newcommand{\mko}{$_{\mathrm MKO}$}
\newcommand{\tmass}{$_{\mathrm 2M}$}
\newcommand{\kms}{km\,s$^{-1}$}

\slugcomment{Accepted to appear in {\it The Astrophysical Journal}, 20 Sep 2012.}
\shortauthors{Tinney et al.}
\shorttitle{A Methane Imaging Y dwarf}

\usepackage{color}

\begin{document}

\title{WISE\,J163940.83-684738.6: A Y dwarf identified by Methane Imaging\altaffilmark{*}}

\author{
C.\ G.\ Tinney\altaffilmark{a},
Jacqueline K. Faherty\altaffilmark{b},
J.\ Davy Kirkpatrick\altaffilmark{c},
Edward L.\ Wright\altaffilmark{d},
Christopher R.\ Gelino\altaffilmark{c},
Michael C.\ Cushing\altaffilmark{e},
Roger L.\ Griffith\altaffilmark{c},
Graeme Salter\altaffilmark{a}
}

\altaffiltext{*}{This paper includes data gathered with the 6.5 meter Magellan Telescopes located at Las Campanas Observatory, Chile.}
\altaffiltext{a}{Department of Astrophysics, School of Physics, University of New South Wales, NSW 2052, Australia, c.tinney@unsw.edu.au}
\altaffiltext{b}{Department of Astronomy, University of Chile, Camino El Observatorio 1515, Casilla 36-D, Santiago, Chile}
\altaffiltext{c}{Infrared Processing and Analysis Center, MS 100-22, California, Institute of Technology, Pasadena, CA 91125}
\altaffiltext{d}{Department of Physics and Astronomy, UCLA, Los Angeles, CA 90095-1547}
\altaffiltext{e}{Department of Physics and Astronomy, MS 111, University of Toledo, 2801 W. Bancroft St., Toledo, OH 43606-3328}

\begin{abstract}

We have used methane imaging techniques to identify the near-infrared counterpart of the bright WISE source WISE\,J163940.83-684738.6. The large proper motion of this source ($\approx$3.0\arcsec\,yr$^{-1}$ ) has moved it, since its original WISE identification, very close to a much brighter background star -- it currently lies within 1.5\arcsec\ of the J=14.90$\pm$0.04 star 2MASS\,16394085-6847446. Observations in good seeing conditions using methane sensitive filters in the near-infrared J-band with the FourStar instrument on the Magellan 6.5m Baade telescope, however, have enabled us to detect a near-infrared counterpart. We have defined a photometric system for use with the FourStar J2 and J3 filters, and this photometry indicates strong methane absorption, which unequivocally identifies it as the source of the WISE flux. Using these imaging observations we were then able to steer this object down the slit of the FIRE spectrograph on a night of 0.6\arcsec\ seeing, and so obtain near-infrared spectroscopy confirming a Y0-Y0.5 spectral type. This is in line with the object's near-infrared-to-WISE J3--W2 colour. Preliminary astrometry using both WISE and FourStar data indicates a distance of 5.0$\pm$0.5\,pc and a substantial tangential velocity of 73$\pm$8\,\kms. WISE\,J163940.83-684738.6 is the brightest confirmed Y dwarf in the WISE W2 passband and its distance measurement places it amongst the lowest luminosity sources  detected to date.
\end{abstract}

\keywords{Brown dwarfs: individual: WISE J163940.83-684738.6; Techniques: photometric; Methods: observational; Parallaxes }

\section{Introduction}

Data from the NASA Wide-field Infrared Survey Explorer \citep[WISE;][]{wright2010} have delivered  unprecedented advances in our understanding of the properties and space densities of the coldest compact astrophysical sources identified outside our Solar System -- the Y-type brown dwarfs\citep{cushing2011,kirkpatrick2012}.

These very cold brown dwarfs have scientific impacts that span multiple astronomical arenas. In the field of star formation, they can deliver a historical record of the star formation process at very low masses and at epochs  billions of years prior to the star forming regions we observe today. In the field of planetary atmospheric theory, they represent low-temperature atmospheres that can be readily observed without the contaminating glare of a host star, and without the photochemical complications introduced by host star irradiation. While in the field of exoplanet searches, they provide nearby, low-luminosity search targets potentially hosting planetary systems of their own.

WISE is readily able to identify these very cold brown dwarfs by their mid-infrared methane absorption bands. 
The shortest wavelength WISE band (hereafter W1) has a central wavelength of 3.4\,$\mu$m, which sits in the 
middle of the strong fundamental methane absorption band near 3.3\,$\mu$m. The second shortest WISE band (hereafter W2), 
has a central wavelength of 4.6\,$\mu$m, where the photosphere is reasonably transparent, so it detects flux from deeper, 
hotter layers in the brown dwarf. As a result, cold brown dwarfs can be identified via their very red W1--W2 colour. 
Thirteen Y dwarfs have been identified and spectroscopically confirmed to date by the WISE Brown Dwarf
Science Team \citep{cushing2011,kirkpatrick2012}.

\section{Observations of WISE\,J163940.83-684738.6}

The WISE source WISE\,J163940.83-684738.6 (hereafter W1639) passes all the selection criteria for being a cold Y-type brown dwarf \citep[see][for details]{kirkpatrick2012}, but the identification of a near-infrared counterpart to the WISE flux has proved difficult. With W2$ = 13.64\pm0.05$ and W1--W2 $>4.24$, W1639 would represent  (if confirmed as a brown dwarf) the brightest Y dwarf in the WISE survey. 

\begin{figure}
   \begin{center}
   \includegraphics[clip=true,width=7.5cm]{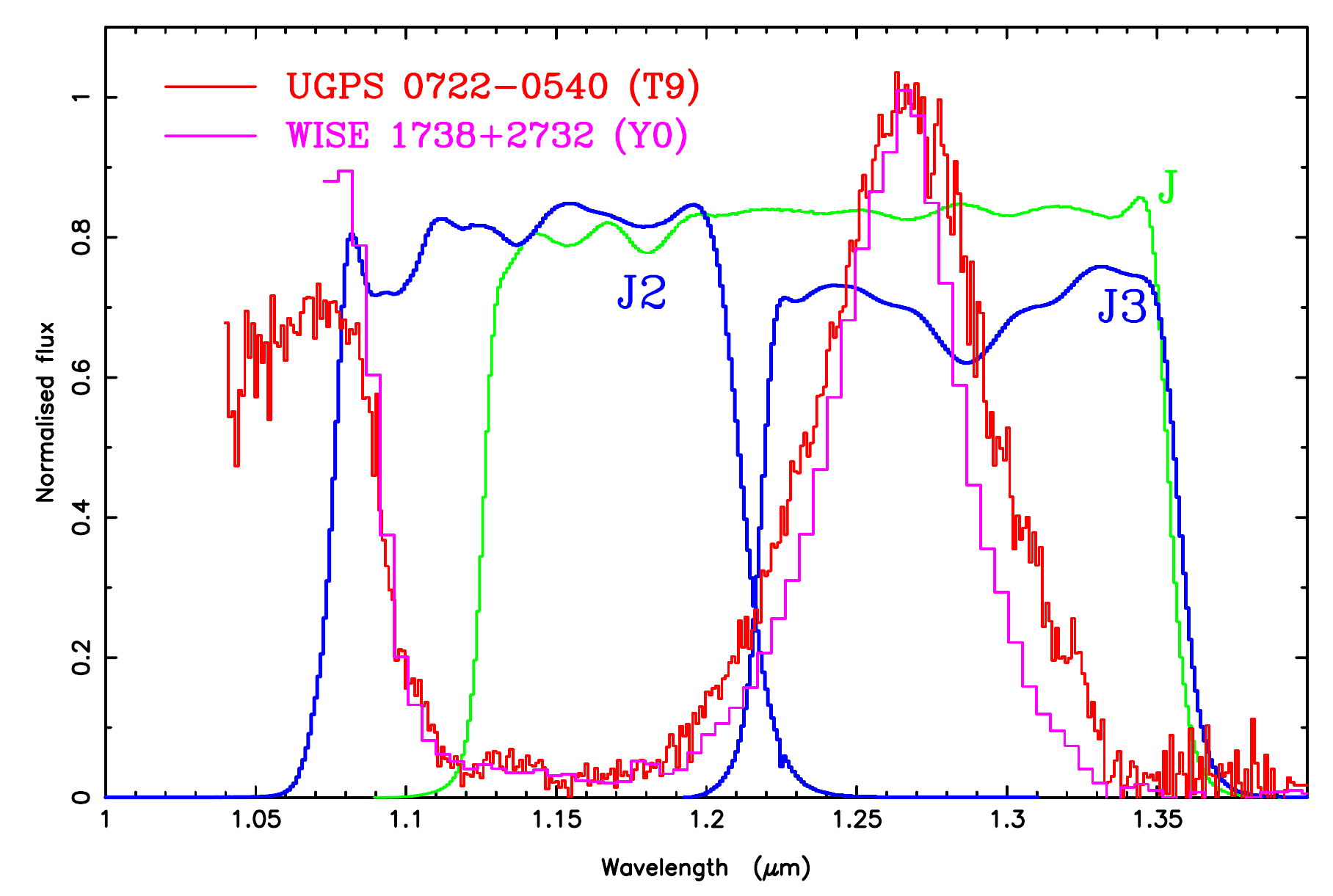}\\
   \end{center}
   \caption{FourStar J, J2 and J3 filter bandpasses, plotted with example very cool brown dwarf spectra (WISEPA\,J173835.53+273258.9 from \citet{cushing2011} and UGPS\,J072227.51$?$054031.2 from \citet{lucas2010} -- the spectral types are those assigned by \citet{kirkpatrick2012}). The J3 filter sees
   significant flux from brown dwarfs with strong methane absorption, while the J2 flux is highly suppressed. }
   \label{bandpasses}
\end{figure}

We have been carrying out observations of WISE sources since early 2012 using the FourStar infrared mosaic camera \citep{persson2008} mounted on the 6.5m Magellan Baade telescope at Las Campanas Observatory, Chile. This program has dual goals: to contribute to the complete identification of all WISE sources with W1--W2$>$2.0; and to carry out astrometric measurements of confirmed cold brown dwarfs. The first of these science goals is pursued using the near-infrared methane-sensitive J2 and J3 filters installed in FourStar. The J2 filter (see Fig. \ref{bandpasses}) is centred on a region in which methane absorption produces strong suppression in very cool brown dwarfs,  while the J3 filter is centred on one of the few opacity holes (at 1.27\,$\mu$m) in cold brown dwarf photospheres. In this respect they represent J-band analogues of the H-band CH$_4$s and CH$_4$l filters used in the methane system of \cite{tinney2005}.

\begin{figure}
   \begin{center}
   (a)\includegraphics[clip=true,width=4cm]{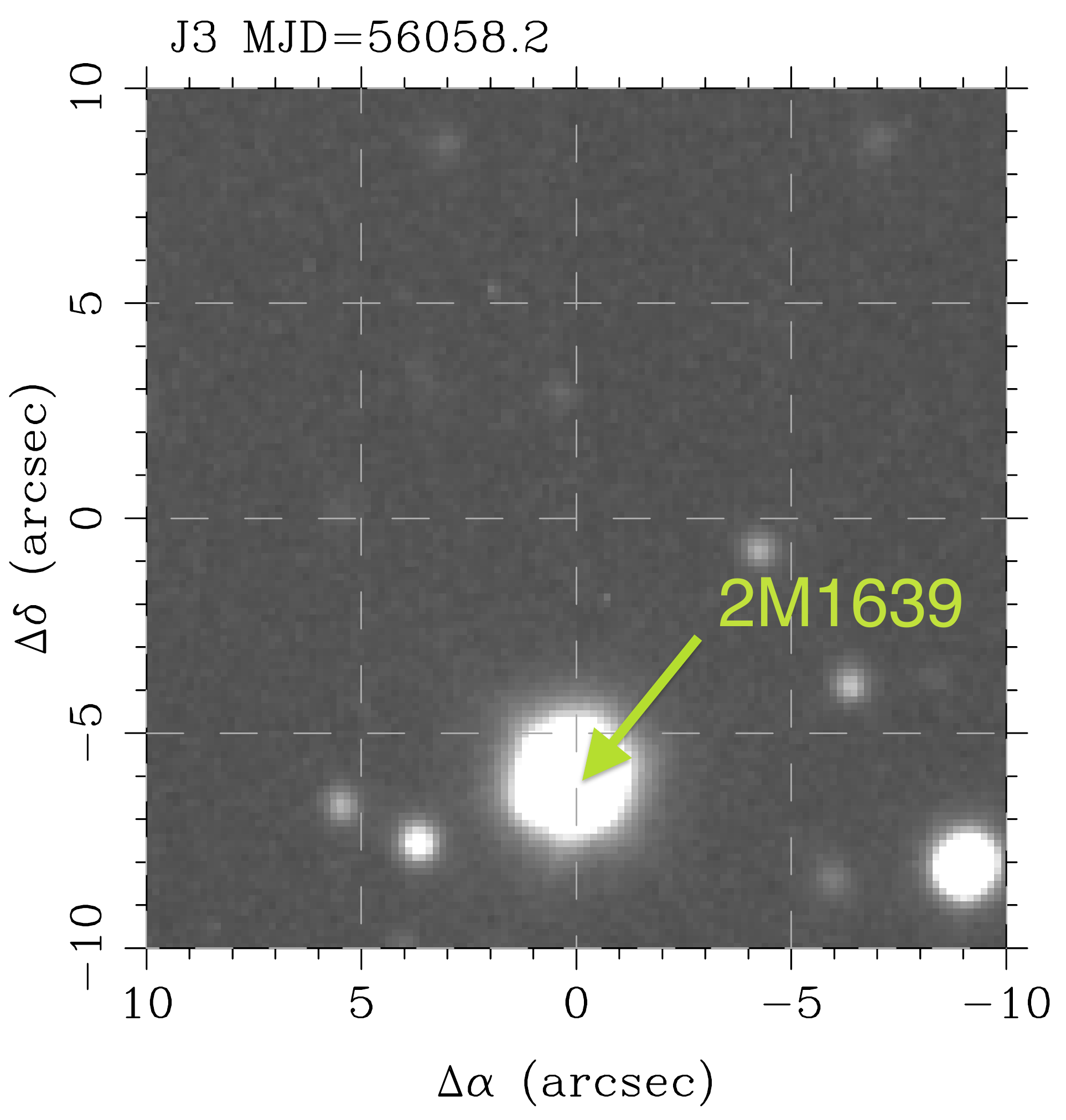}(b)\includegraphics[clip=true,width=4cm]{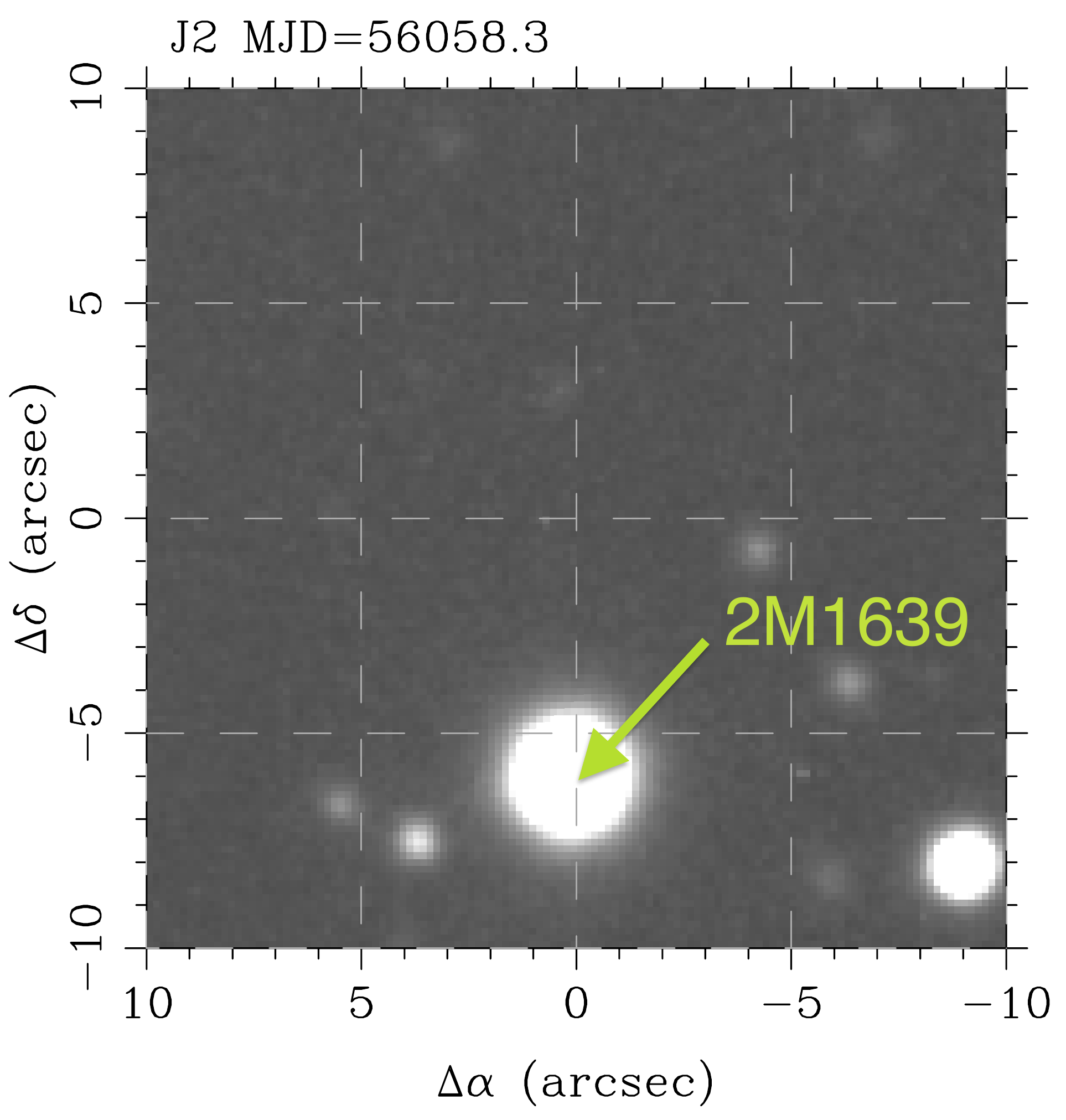}\\[2mm]
   (c)\includegraphics[clip=true,width=4cm]{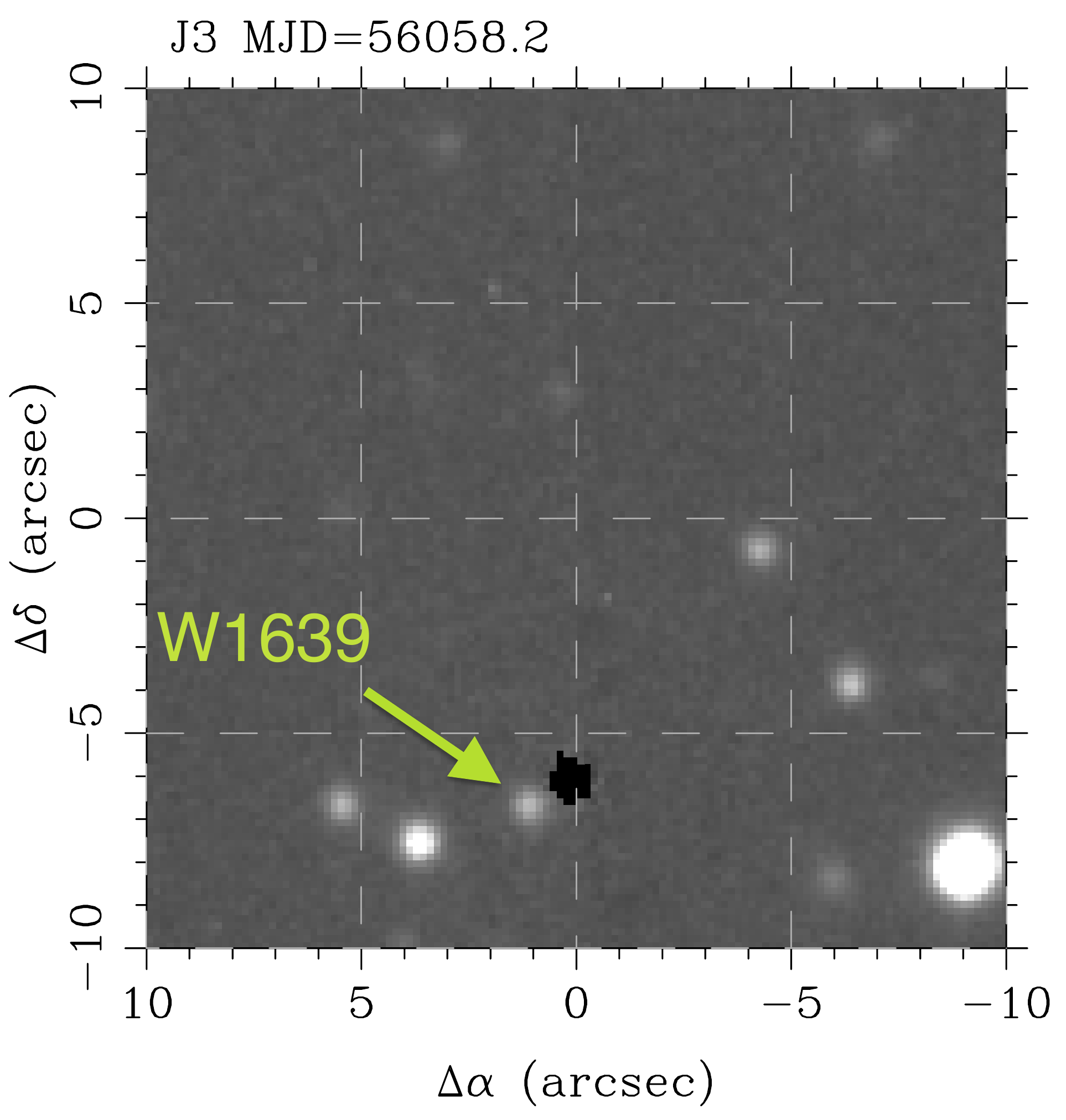}(d)\includegraphics[clip=true,width=4cm]{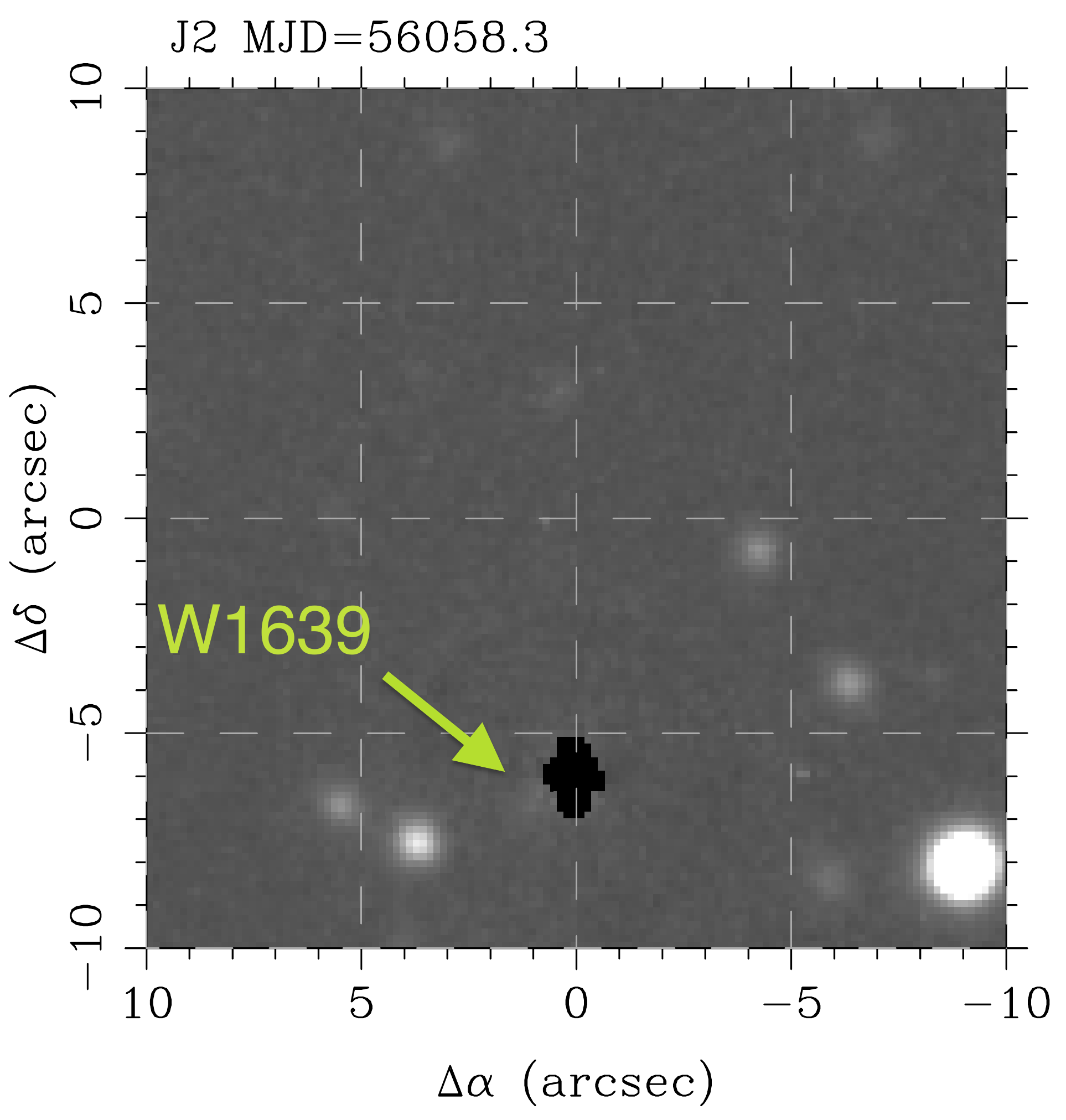}\\[2mm]
   \end{center}
   \caption{FourStar imaging data centred on the W1639 {\em WISE All Sky} release position, taken in the J2 and J3 filters. Panels (a) and (b) show the data from our standard processing, and panels (c) and (d) show the images with the bright source 2M1639 subtracted (as described in the text). The black regions in the lower panels are flagged as bad due to the excess noise arising from imperfect PSF subtraction in 2M1639's core. All panels are plotted with north to the top, and west to the right.}
   \label{charts}
\end{figure}

W1639 was observed on 2012 May 10-11 (UT)\footnote{All dates and times used in this paper are UT dates and times.} with FourStar for approximately 4h, using 50 exposures of 120s in each filter and random telescope dithering. Each of FourStar's four 2048$\times$2048\,pixel detectors sees a field 325\arcsec\ on a side -- in this analysis we deal only with data from a single detector {\em (Chip2)} in which all our targets were placed.
Data processing was performed using a version of the ORACDR pipeline system \citep{cavanagh2008}\footnote{See also \tt\url{http://www.oracdr.org/}} modified by us for use with FourStar. This creates and subtracts averaged dark frames, combines observations of a given target in each filter to create a flat-field, flattens all frames and then mosaics them to produce a final image. The results of each first-pass processing with ORACDR were used to identify frames with poor image quality and/or poor transmission, and these were then removed before a second-pass reprocessing. The final images for W1639 are created from the 29 best J3 images (totalling 58\,min of exposure time and delivering 0.66\,\arcsec\ images) and the 27 best J2 images (totalling 54\,min and delivering 0.81\,\arcsec\ images). The resulting images, centred on W1639's {\em WISE All Sky} release position, are shown in Figure \ref{charts}(a) and (b). No obvious source is visible with the very red J3--J2 colour expected for a cool brown dwarf. The bright source to the south of W1639's position is the J=14.90$\pm$0.04 star 2MASS\,J163940.85-684744.6 (hereafter 2M1639).

These images were then photometered using the DAOPHOT II package \citep{stetson1987}, as implemented within the Starlink environment \citep{draper2005}\footnote{See also \tt\url{http://starlink.jach.hawaii.edu/starlink}}. Unsaturated point-spread function (PSF) stars were selected within a 1' radius of W1639, and used to create an initial model PSF. This was used to fit and subtract all identified stars within the image. In many cases this reveals the presence of companion stars within the haloes of PSF stars. Identifying these stars and including them in the PSF analysis on a second pass allowed us to iterate towards an ``uncontaminated'' PSF, which was then used to fit and subtract the bright star 2M1639. The resulting images ``cleaned'' of this bright star are shown in Figure \ref{charts}(c) and (d), and demonstrate the existence of a near-infrared source previously hidden within the halo of 2M1639. Moreover this source is bright in J3 and faint in J2 -- precisely the signature expected if the WISE W1639 flux is due to a cool brown dwarf.

Subsequently, the same field was re-observed as an astrometric target in the J3 filter for around one hour in each of the nights 2012 July 10,11 and August 10. This data is discussed further in \S5.

\section{Photometry on the J3/J2 system.}

We can use imaging data to ask what the J3--W2 colours for W1639 imply for this system. We have used SExtractor \citep{bertin96} to obtain aperture photometry for stellar sources in both the ``cleaned'' W1639 images (obtained as described above), and for processed images of other late-T and Y dwarfs observed with FourStar as part of our WISE follow-up program on the nights of 2012 March 9, May 10 and July 10-11. These uncalibrated photometric catalogues were then position-matched against 2MASS catalogue data in these fields for both astrometric and photometric calibration.

\begin{deluxetable*}{lcclllllll}
\tabletypesize{\scriptsize}
\tablenum{1}
\tablecaption{FourStar J3/J2 photometry for W1639 and other WISE T and Y dwarfs.}
\tablewidth{18cm}
\tablehead{
\colhead{Object}          & \colhead{Type} 
                                  & \colhead{Ref\tablenotemark{a}}   
                                        & \colhead{J\mko\tablenotemark{b}}
                                                        & \colhead{Obs.Date}
                                                                                & \colhead{J3\tablenotemark{c}}
                                                                                                   & \colhead{J2}
                                                                                                                   & \colhead{J3-J2}
                                                                                             	                                     &\colhead{W2}
	         																				                                                                             &\colhead{J3--W2}
}
\startdata
WISEPC J014807.25-720258.7 & T9.5 & 1,5   & 18.96\,0.07 & Jul\,6                & 19.83\,0.09    & 20.57\,0.10 & $-$0.74\,0.14 & 14.69\,0.05 & 5.14\,0.13 \\
WISE J053516.80-750024.9   & $\ge$Y1:  & 1 & \nodata        & Mar\,9                & 21.98\,0.13    & \nodata        & \nodata          & 15.06\,0.07 & 6.92\,0.15 \\
WISE J071322.55-291751.9   & Y0   &  1  & 19.64\,0.15 & Mar\,9                & 19.48\,0.03    & 21.00\,0.05 & $-$1.52\,0.06 & 14.48\,0.06 & 5.00\,0.07 \\
WISE J073444.02-715744.0   & Y0   &  1  & 20.41\,0.27 & Mar\,9                & 20.06\,0.04    & \nodata        & \nodata          & 15.36\,0.06 & 4.70\,0.07 \\
WISE J081117.81-805141.3   & T9.5 &  2  & \nodata        & Mar\,9,May\,10        & 19.32\,0.03 (2)& 21.04\,0.09 & $-$1.72\,0.10 & 14.38\,0.04 & 4.95\,0.05\\
WISE J091408.96-345941.5   & T5   &  3  & \nodata        & May\,10               & 17.73\,0.02    & 19.03\,0.02 & $-$1.30\,0.03 & 15.03\,0.09 & 2.70\,0.09 \\
WISEPC J104245.23-384238.3 & T8.5 &  4  & \nodata        & Mar\,9,May\,10,Jul\,7 & 18.57\,0.02 (3)& \nodata        & \nodata          & 14.52\,0.06 & 4.05\,0.06 \\
WISE J150115.92-400418.4   & T6   &  3  & \nodata        & May\,10               & 15.95\,0.01    & 17.01\,0.01 & $-$1.06\,0.02 & 14.21\,0.05 & 1.74\,0.05 \\
WISEPA J154151.66-225025.2 & Y0.5 &  1,6  & 20.74\,0.31 & May\,10,Jul\,7        & 20.96\,0.05 (2)& 21.48\,0.10\tablenotemark{d} 
                                                                                                                      & $-$0.52\,0.10\tablenotemark{d}
                                                                                                                                & 14.25\,0.06 & 6.77\,0.09  \\
WISE J163940.86-684744.6   & Y0:  &  6  & \nodata        & May\,10               & 20.62\,0.08    & 22.27\,0.10 & $-$1.65\,0.12 & 13.64\,0.05 & 6.98\,0.09 \\
WISE J210200.15-442919.5   & T9   &  3  & \nodata        & May\,10,Jul\,6,7      & 18.06\,0.02 (3)& 19.38\,0.09 & \nodata          & 14.12\,0.05 & 3.93\,0.06 \\
WISEPA J213456.73-713743.6 & T9p  &  1  &                & Jul\,6,Jul\,7        & 19.38\,0.09    & \nodata        & \nodata        & 13.99\,0.05 &  5.39\,0.06 \\                      
WISE J222055.31-362817.4   & Y0   &  1  & 20.38\,0.17 & May\,10               & 20.10\,0.05 (2)& 21.75\,0.20 & $-$1.65\,0.2  & 14.66\,0.06 & 5.45\,0.09 \\
\enddata
\label{photometry}
\tablenotetext{a}{Spectral Type Sources are: 1 -- \cite{kirkpatrick2012}; 2 -- Mace et al. in prep.; 3 -- Tinney et al. in prep,; 4 -- \cite{kirkpatrick2011}; 5 -- \cite{cushing2011}; 6 -- this paper.}
\tablenotetext{b}{J\mko\ values from \cite{kirkpatrick2012}}
\tablenotetext{c}{Where multiple measures are available, the mean is reported along with the number of observations averaged.}
\tablenotetext{d}{J2 photometry poor due to contaminating J2=18.6 star 1.3\arcsec\ away. The real uncertainty will be much larger than the formal photon-counting uncertainty reported here.}
\end{deluxetable*}

With no pre-existing standards to define a J2/J3 photometric system, we are forced to define our own system \cite[as we did in][]{tinney2005}. Following that example, we base our system on Mauna Kea Observatories (MKO) J-band photometry -- i.e. we adopt J\mko\ standards over a narrow colour range to define our magnitude zero-point for J3 and J2. We obtain J\mko\ photometry for stars in each field by converting 2MASS JHK$_{\mathrm s}$ photometry to the MKO system, for stars in the colour ranges where the transformations in the {\em Explanatory Supplement to the 2MASS All Sky Data Release and Extended Mission Products}\footnote{http://www.ipac.caltech.edu/2mass/releases/allsky/doc/} are reliable (i.e. in the range $-0.2 < $(J--K$_{\mathrm s}$)\tmass$ < 1.5$ and $-0.2 < $(J--H)\tmass$ < 1.1$). 

\begin{figure}
   \begin{center}
   \includegraphics[clip=true,width=8.5cm]{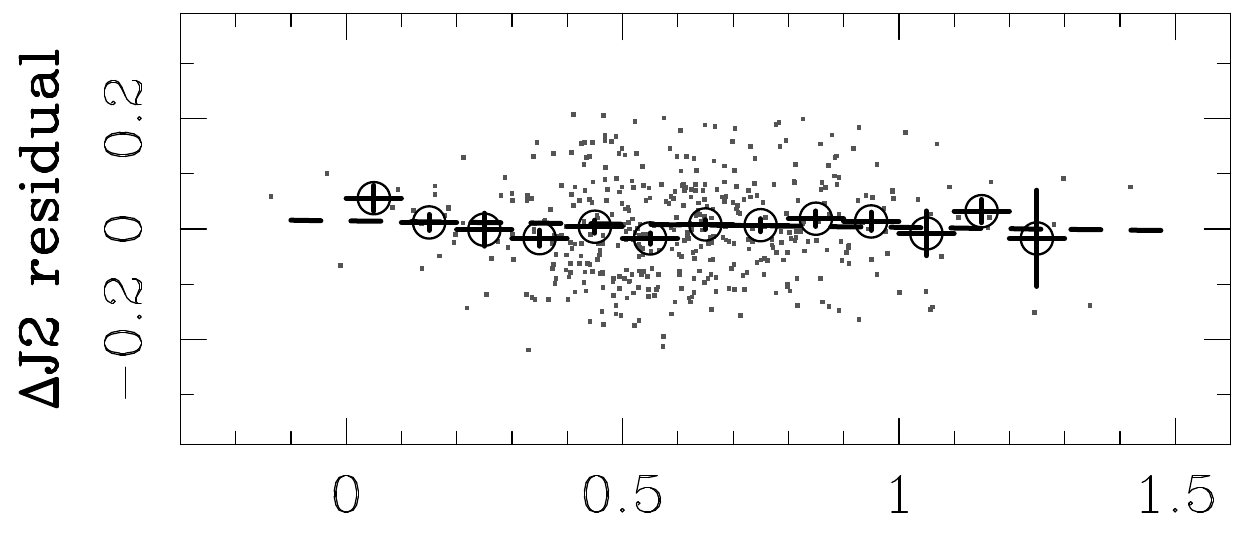}\\[2mm]
   \includegraphics[clip=true,width=8.5cm]{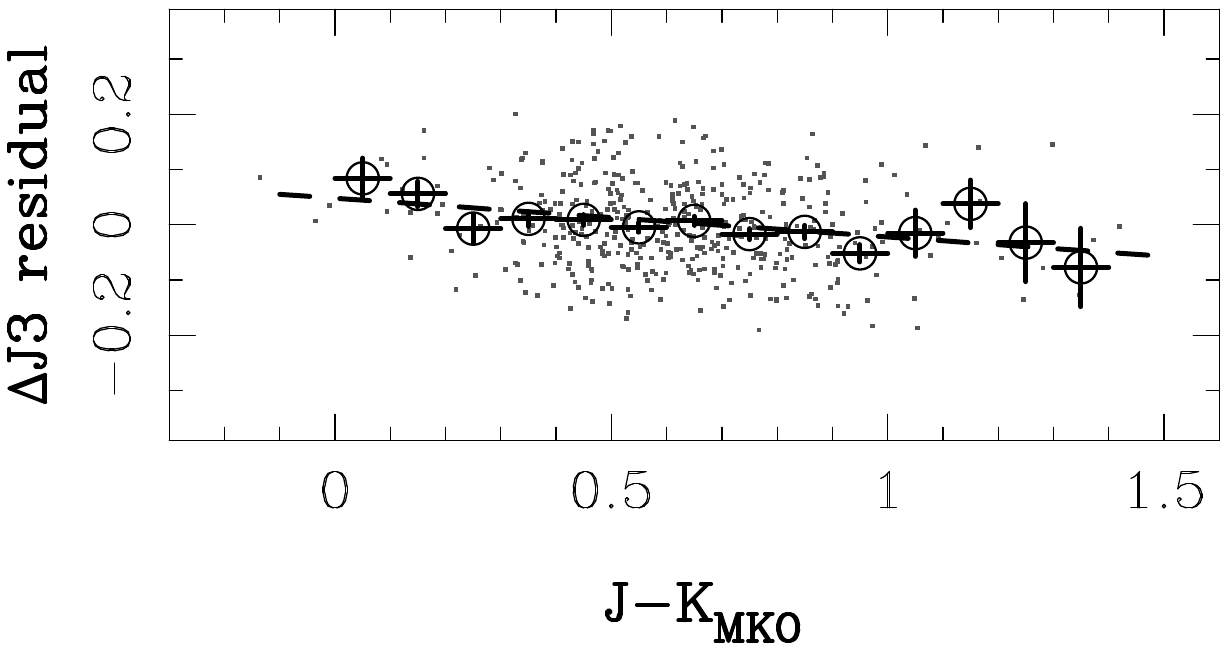}\\
   \end{center}
   \caption{Residuals about J2 and J3 zero-point calibrations (defined by objects with $0.4 < $ J--K\mko\ $ < 0.8$) as a function of J--K\mko. Over 400 stars are plotted as small dots in each panel, representing combined results from nine FourStar fields with one detector. Large circles show the binned means of the dots, while the dashed lines show linear fits to the binned data.}
   \label{colourterms}
\end{figure}

In \citet{tinney2005} we used relatively hot A,F and G-type dwarf stars to define a CH$_4$s and CH$_4$l magnitude system in the H-band. Unfortunately, such stars (with colours in the range $-0.02 < $ (J--K)\mko\ $ < 0.50$) are rare in an arbitrary FourStar field. Instead we base our J2/J3 system on stars in the range $0.4 < $ (J--K)\mko\ $ < 0.8$, corresponding to the spectral type range G6-M3. Fortunately the colour terms involved in the calibration of this J2 and J3 photometry using J\mko\ are small. Figure \ref{colourterms} shows a plot of the residuals about this calibration, as a function of (J--K)\mko\ for more than 400 stars in nine FourStar Chip2 fields -- the small dots show individual data points, and the large circles show the results of binning these points into 0.1\,mag bins in colour, while  the dashed line shows a linear fit. There is no evidence for systematic colour terms in the residuals for J2 over the range $0.0 < $ (J--K)\mko\  $ < 1.3$. The colour term present in J3 over this range is small ($\approx$0.07\,mag$/$mag) meaning that using stars with a mean J--K of 0.6, rather than 0.25, would at worst introduce a potential zero-point error of 0.025\,mag. This is at a level similar to the standard-error-in-the-mean for the zero-points obtained in each field, and so neglecting the colour-term is acceptable.

\begin{figure}
   \begin{center}
   \includegraphics[clip=true,width=6cm]{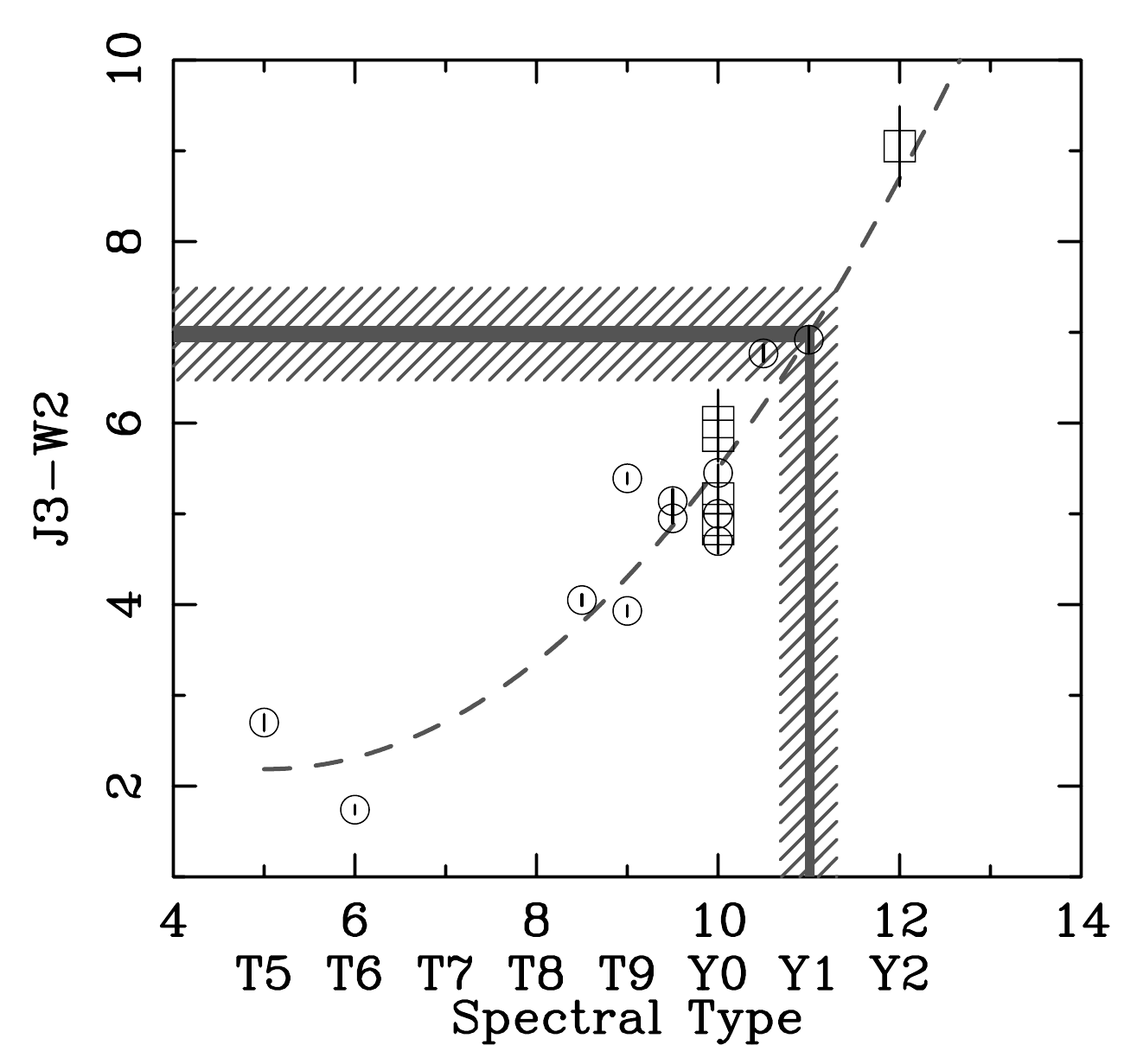}\\
   \end{center}
   \caption{Relationship between J3-W2 and spectral type. Circles are the photometry in Table \ref{photometry}, and  squares are Y dwarfs with J\mko\ photometry from \cite{kirkpatrick2012} converted to J3 as described in the text. The dashed line is a quadratic polynomial fit (J3--W3 = $5.55 - 1.3428\,n + 0.13373341\,n^2$,  where $n$ is the numeric parametrisation of the T/Y type) with a scatter about the fit of $\pm$0.50 in J3-W2. The photometry of W1639 maps it to an equivalent spectral type of Y1 (solid shaded region), while the scatter about the fit (hatched region) maps onto an uncertainty of $\pm$0.3 sub-types.}
   \label{J3mW2vSpT}
\end{figure}

The result is photometry on the natural system of these J2 and J3 filters, with a zero-point measured using J\mko\ photometry for stars with $0.4 < $ (J--K)\mko\ $ < 0.8$. This system has the significant advantage that it can deliver high quality differential J2/J3 photometry without the need for absolute photometric conditions. The FourStar field of view typically delivers between 5 and 100 2MASS stars suitable for calibrating each FourStar detector, and typical uncertainties on the determination of this zero-point are in the range 0.01-0.02\,mag. Table \ref{photometry} presents photometry on this system for both W1639 and a sample of comparison late-T and Y dwarfs observed with FourStar. Also listed are the relevant {\em WISE All Sky} release photometry and spectral classifications.

Figure \ref{J3mW2vSpT} shows J3--W2 colours from Table \ref{photometry} as a function of spectral type. The Y dwarfs in Table \ref{photometry} with both J\mko\ and J3 photometry indicate a mean offset between these two systems of J\mko$-$J3=0.14$\pm$0.25 for Y dwarfs\footnote{Essentially all the Y dwarf flux in the J band emerges within the J3 bandpass (see Fig.1), so a simple correction is appropriate -- at present most of the scatter in this zero-point is due to photon-counting uncertainties for these faint targets.} Applying this correction to the published J\mko\ photometry for WISE\,J173835.53+273259.0, J140518.39+553421.3, J035934.06$-$540154.6, J041022.71+150248.4, J182831.08+265037.7 and J205628.91+145953.2 in \citet{kirkpatrick2012} allows us to add  computed J3--W2 colours and spectral types for these objects (plotted with square symbols) to Fig. \ref{J3mW2vSpT}.

Much like the plots of J\mko--W2 presented in \citet{kirkpatrick2011}, it is clear that J3--W2 provides considerable leverage on the measurement of how cool  a late-T or Y dwarf is -- particularly when combined with a J3--J2 colour. The latter allows one to determine whether an object has significant methane absorption (i.e. is J3--J2 $\lesssim -0.5$), in which case J3--W2 can be relied on to be a monotonic estimator of how cool the brown dwarf is (or equivalently of its spectral type). In the case of W1639 (which has almost exactly the same J3--W2 as the $\ge$Y1: brown dwarf WISE J053516.80+750024.9), this procedure produces an spectral type estimate of Y1 with an uncertainty of $\pm$0.3 sub-types (arising from the scatter about the calibration curve in Fig. \ref{J3mW2vSpT} rather than the precision of the J3 photometry). 

\section{FIRE Spectroscopy}

Spectroscopic observations of W1639 were carried out on 2012 July 10 (UT) using the Folded-port Infrared Echellette \citep[FIRE;][]{simcoe2008,simcoe2010} on the Magellan Baade Telescope. FIRE uses a 2048$\times$2048 HAWAII-2RG array and in prism mode delivers a wavelength range from 0.8 to 2.5$\mu$m at a resolution of $\lambda/\Delta\lambda$ $\approx$ 500 in the J band (when used with a 0.6\arcsec\ slit). 

Our FourStar astrometric observations on the night of 2012 July 10 were processed on-the-fly using ORACDR, to obtain a precise position angle and separation at this epoch between the bright source 2M1639 and the faint source W1639. FIRE was rotated to this position angle, and 2M1639  positioned on the slit, ensuring that W1639 would also be acquired down the slit. Six 600s exposures were obtained nodding the slit by 10\arcsec\ in an ABBA pattern.

These spectra were then processed in a standard fashion, using arc lines to remove FIRE's optical curvature and traces of the bright star 2M1639 to remove a slight tilt of the spectra with respect to the detector rows. Images were pair subtracted and combined to produce a pair of A-B and B-A images, and these then had any residual sky extracted and subtracted. 

\begin{figure}
   \includegraphics[clip=true,width=9cm]{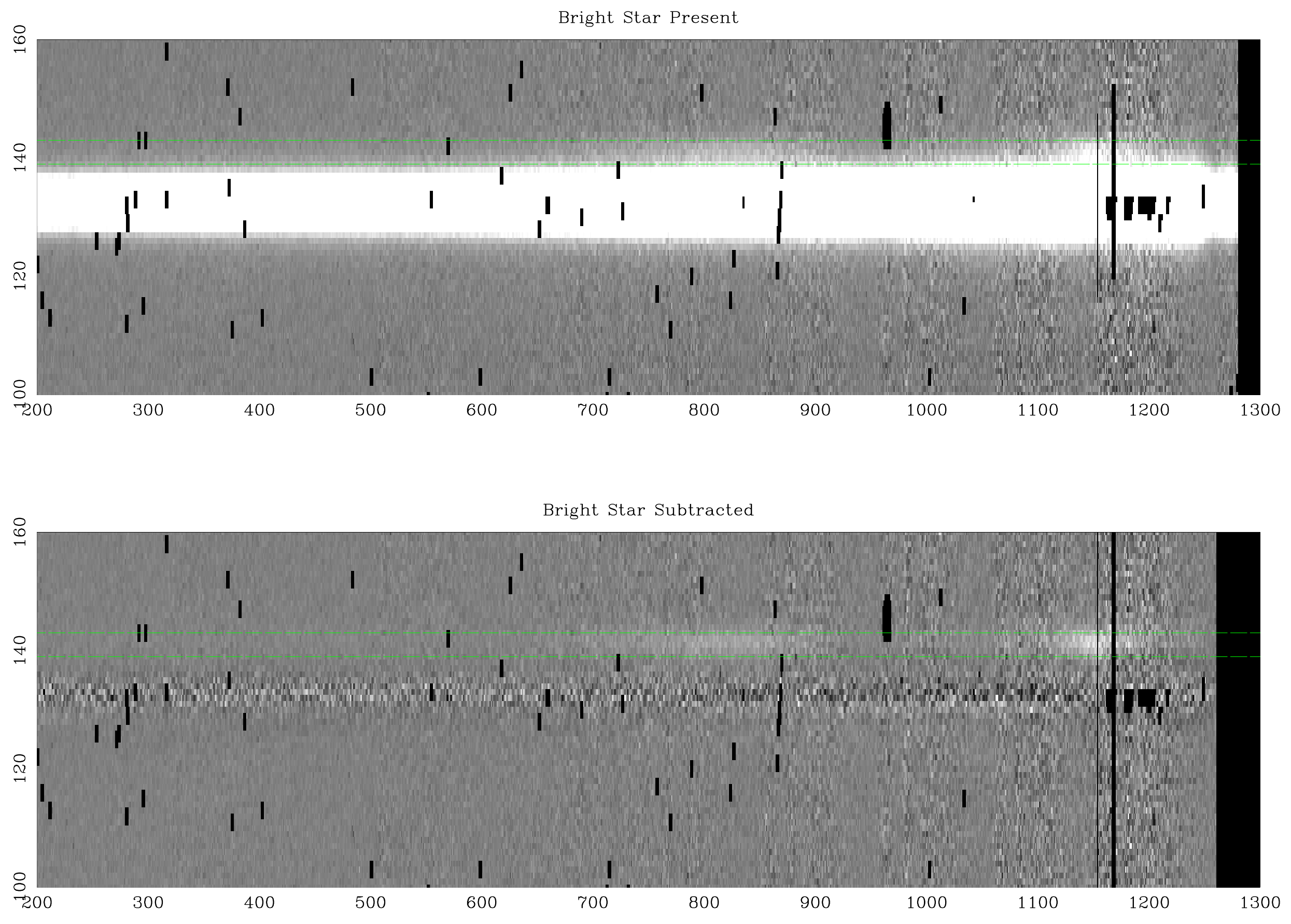}
   \caption{FIRE spectroscopy of 2M1639 and W1639 {(\em upper panel)} observed together along the FIRE slit, and {\em (lower panel)} after the removal of the signature of 2M1639, using the spatial point-spread function procedure described in the text. The overlaid dashed lines show the location of a 0.6" wide extraction region centred on the position of W1639. The axes are labelled in FIRE pixel co-ordinates, after the removal of distortions as described in the text.}
   \label{extraction}
\end{figure}

\begin{figure}
   \includegraphics[clip=true,width=8cm]{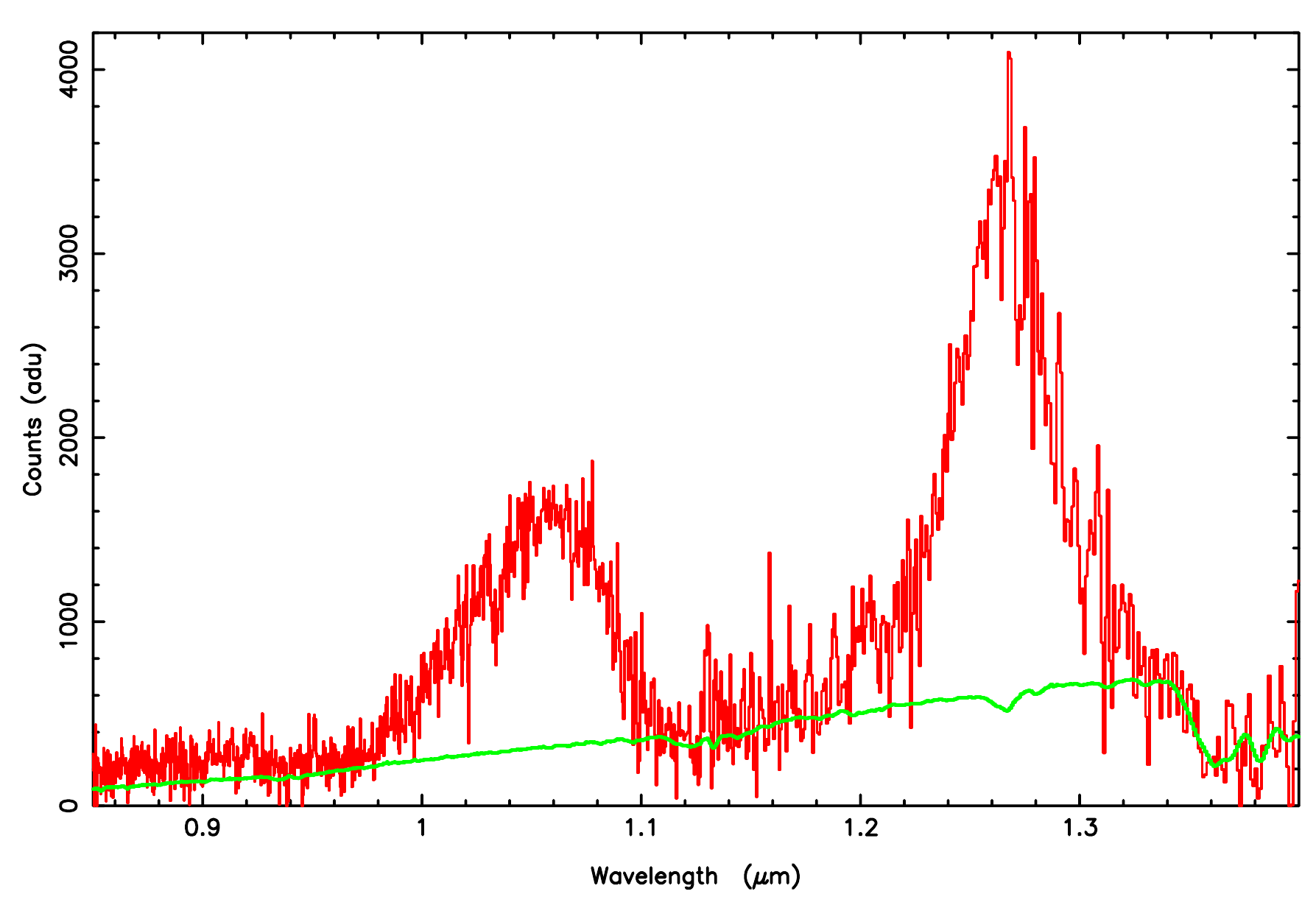}
   \caption{FIRE spectrum of W1639 extracted as described in the text, with a scaled version of the 2M1639 spectrum overplotted to match the flux seen in the contaminated spectrum in the 1.36-1.40$\mu$m range where the flux emitted by a cool brown dwarf will be completely negligible. The vertical scale is in counts per 3600s.}
   \label{extracted}
\end{figure}

The resulting images are dominated by flux from the bright source 2M1639, together with much fainter emission from W1639 (which is offset by +9.2 pixels in the FIRE data). The expected spectrum of a very-late-T or Y dwarf can be seen nonetheless in this data (see Fig. \ref{extraction} - {\em upper panel}). To remove the contaminating flux due to the bright companion, we implemented a ``spectral PSF'' subtraction using custom scripts in the Perl Data Language\footnote{\url{http://pdl.perl.org}}. We extracted spatial profiles from the spectra by averaging every 100 pixels in the spectral direction, and fitting a model PSF to the resulting spatial profile -- after much experimentation we settled on a Gaussian model for the core of the spectral PSF, a Moffat function for its wings, and we also allowed for a ``companion'' Gaussian (with the same width as the core) at the spatial separation known between W1639 and 2M1639, but with an amplitude allowed to float as a free parameter. We found that even this model, however, left systematic residuals when subtracted, and so we followed the lead of DAOPHOT \citep{stetson1987} and allowed for an empirical look-up table of residuals as a fraction of the total peak height of the spectral PSF. The free parameters for the spectral PSF were then fitted with quadratic polynomials to generate a model PSF that varied smoothly with wavelength. This PSF (minus the ``companion'' terms that modelled W1639) was then generated, normalised to each column of the image, and subtracted to create a ``cleaned'' spectral image (see Fig. \ref{extraction} - {\em lower panel}). Spectra were then extracted from this image, wavelength calibrated and averaged over the A-B and B-A beams to produce a final extracted spectrum (see Fig. \ref{extracted}).

\begin{figure}
   \includegraphics[clip=true,width=9cm]{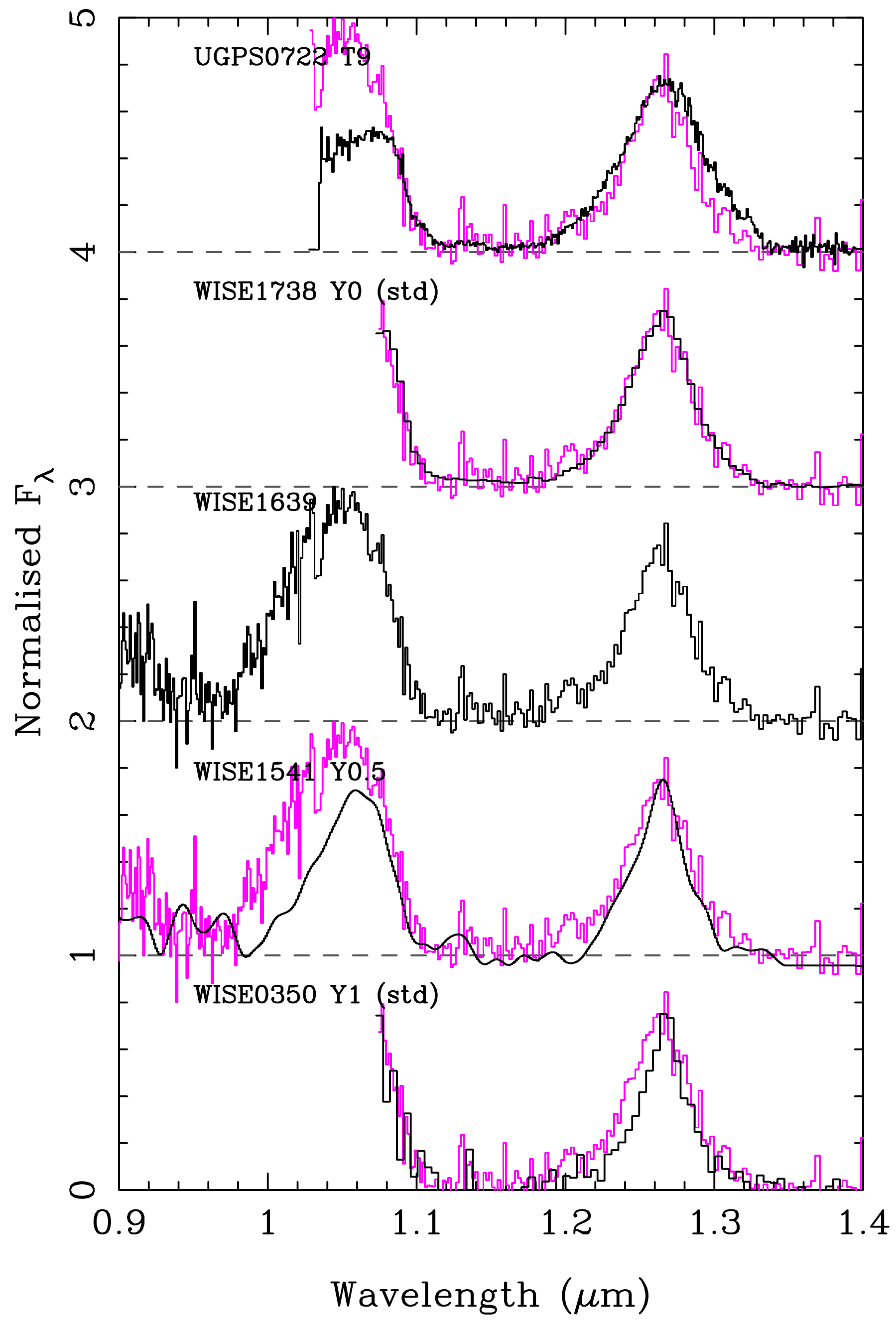}
   \caption{FIRE spectrum of W1639 compared \citep[following][]{kirkpatrick2012} with late-T and Y standards. All spectra have been normalised to value 0.75 at the J-band peak, and had an arbitrary offset applied for display purposes. The W1639 spectrum is shown in black in the central spectrum, and reproduced in magenta (in electronic colour version) or grey (in published paper version) behind the other spectra.}
   \label{types}
\end{figure}

Despite the considerable care with which we removed the 5 magnitudes brighter spectrum of 2M1639, Fig. \ref{extracted} indicates that there is nonetheless some contaminating 2M1639 flux still present in the extracted spectrum. This is demonstrated by the flux seen in the extracted spectrum at 1.36-1.4$\mu$m -- very cool brown dwarfs emit essentially {\em zero} flux at these wavelengths, and they {\em certainly} don't show sharp atmospheric absorption at 1.36$\mu$m. The spectrum of 2M1639 does, however, show such a feature and a scaled version of that spectrum (shown in the figure) is an excellent match to the ``baseline'' flux present in the extracted spectrum. We are therefore confident that removing the scaled 2M1639 spectrum, as shown, is required to generate a decontaminated spectrum of W1639. This decontaminated spectrum was then binned by a factor of 3 (to match spectral pixels to the FIRE resolution in this wavelength range) and a telluric star correction applied to remove atmospheric absorption as well as place our spectra on an F$_\lambda$ scale. The final spectrum is shown in Figure \ref{types}.

We then follow the procedure outlined in \citet{kirkpatrick2012}, and rank the relative lateness of W1639 against other Y dwarfs using a comparison of spectra scaled by the flux peak in the J-band. W1639 is clearly seen to have a J-band peak narrower than that of the T9 prototype UGPS\,0722, and itself has a J-band peak significantly broader than that of the Y1 prototype WISE\,0350. It matches best the Y0 prototype, but given the quality of the spectra we have obtained for this difficult-to-observe object, it could lie anywhere in the range Y0-Y0.5.

\begin{figure}
	\begin{center}
   \includegraphics[clip=true,width=7cm]{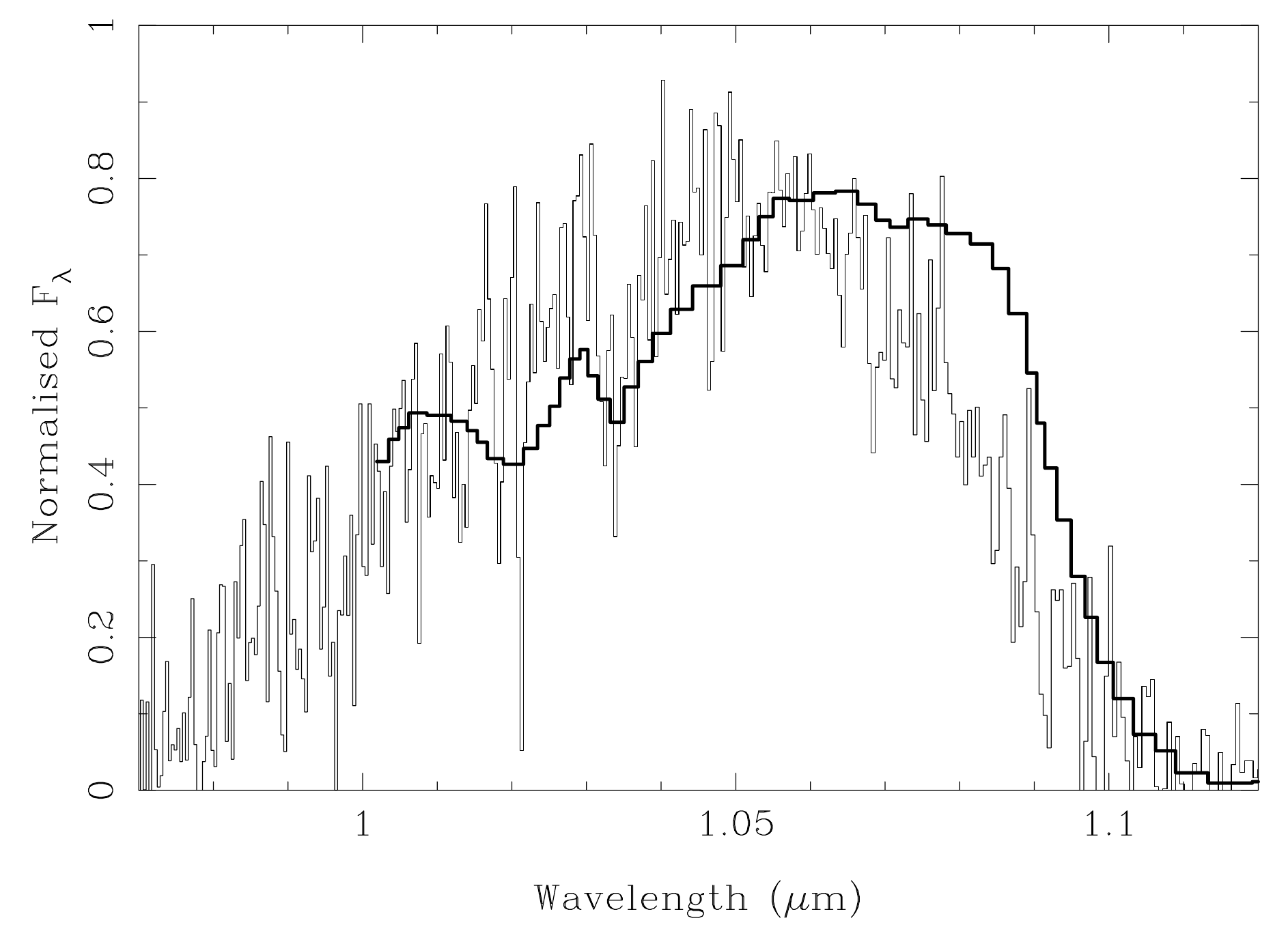}
   \caption{Expanded region of the unbinned spectra in the vicinity of 1.03\,$\mu$m, along with the 500K model of \cite{Saumon2012} including opacities 
   due to NH$_3$ (thick line). The two spectra have been normalised to F$_\lambda$=0.75 in the range 1.045-1.055\,$\mu$m.}
   \label{ammonia}
	\end{center}
\end{figure}

It is tempting to associate the features seen near 1.03\,$\mu$m in this spectrum with the NH$_3$ features near 1.03\,$\mu$m predicted by the models of \citet[][see their Figure 7]{Saumon2012}. To examine this we plot in Fig. \ref{ammonia} the W1639 spectrum near this wavelength before rebinning, along with the 500K BYTe NH$_3$ model in Saumon et al.'s Fig. 7. The alignment of the predicted NH$_3$ ``doublet'' feature near 1.02 and 1.035$\mu$m is tantalising, but due to the noise added to the spectral extraction by the contaminating flux of the bright companion, we are unable to claim this as a robust NH$_3$ detection from this data alone.

\section{Astrometry}

W1639 has now been observed at four epochs with Magellan on three observing runs. It was also observed at two epochs by WISE. Given its very large proper motion, and the fact a Y spectral type and J3=20.62$\pm$0.08 would suggest a likely distance of less than 10\,pc, an astrometric analysis is  warranted.

\begin{figure*}
   \includegraphics[clip=true,height=9.0cm]{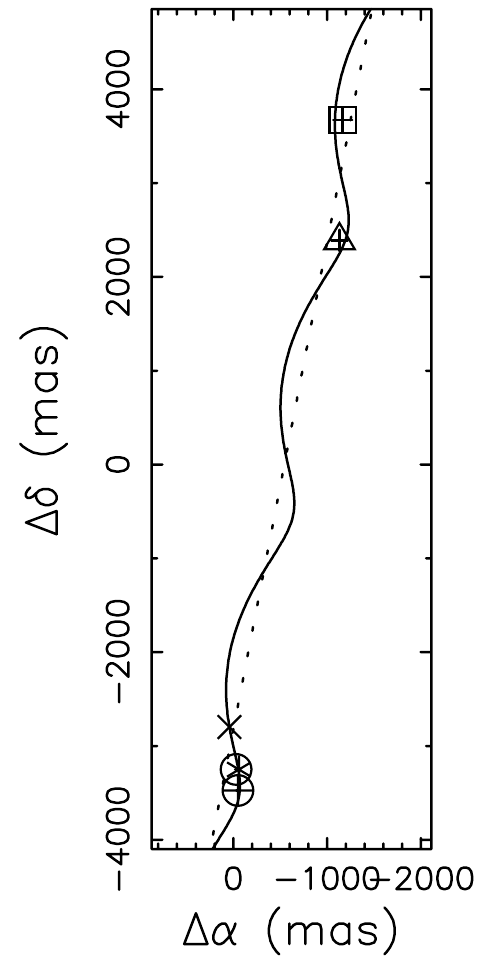}\includegraphics[clip=true,height=6.0cm]{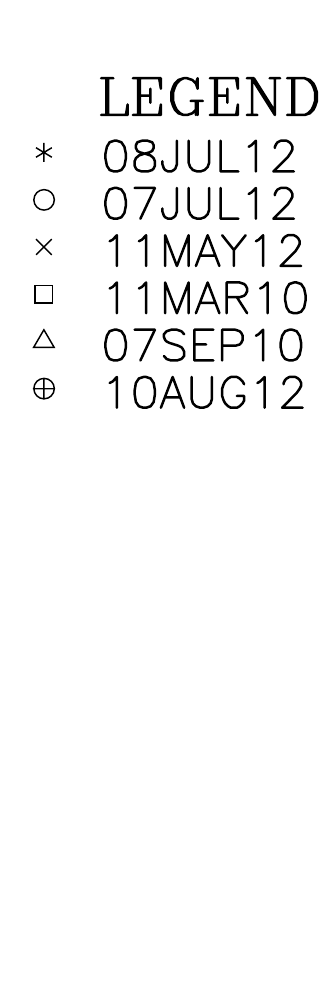}\includegraphics[clip=true,height=9.0cm]{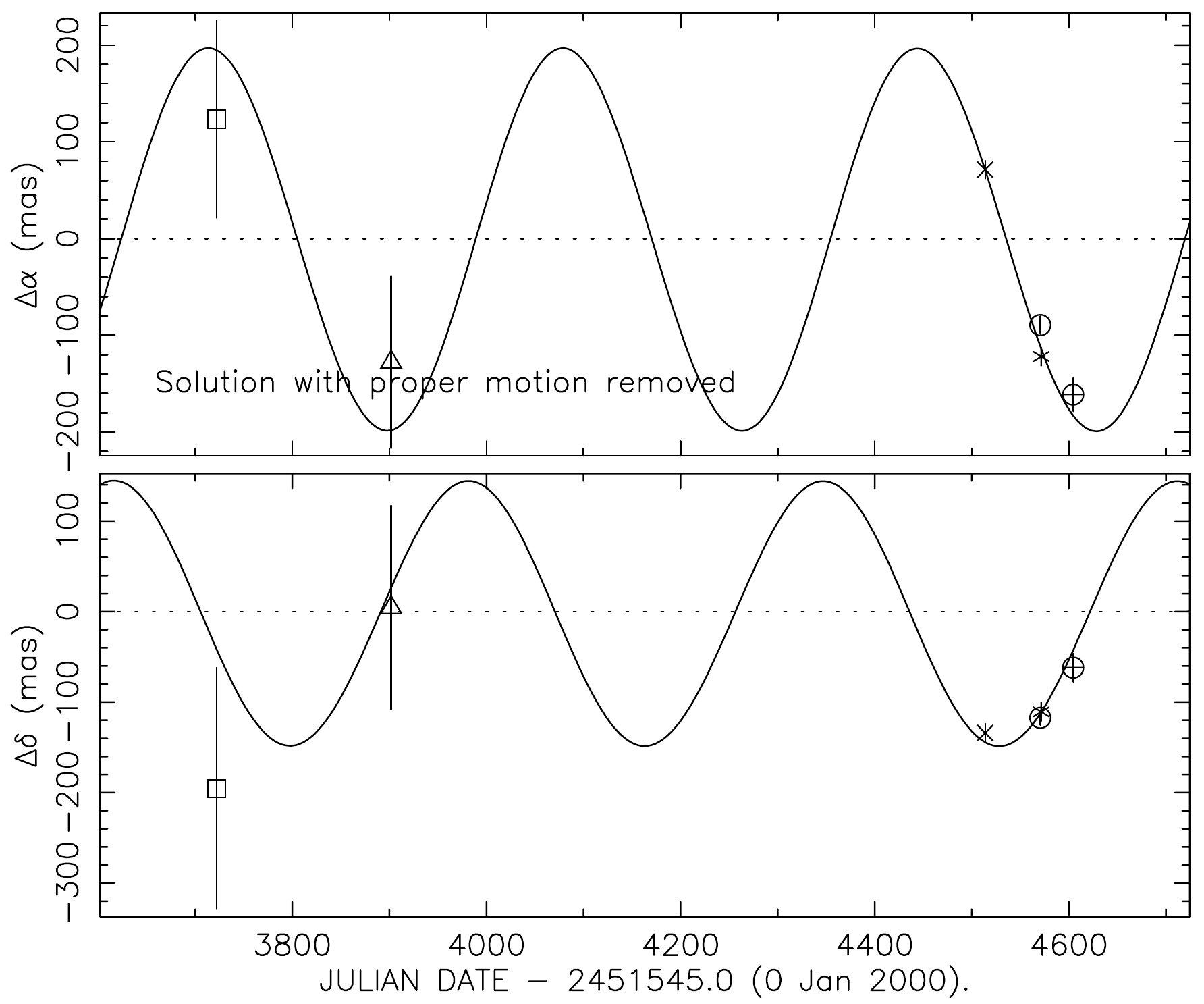}
      \caption{WISE and FourStar astrometry for W1639. {\em Left panel} shows actual motion on the sky with a proper motion and parallax solution superimposed. The WISE data are the two north most points (i.e. at the top of the panel). {\em Right panels} show the right ascension and declination solutions with the fitted proper motion removed for clarity. The WISE data are the two leftmost data points in each panel.}
   \label{astrometry}
\end{figure*}

{\em FourStar : } Our J3-passband, ORACDR-processed images were analysed using DAOPHOT to obtain precision differential astrometry in a manner identical to that used by us in previous astrometric papers \citep{tinney2003,tinney1996}. We use local 2MASS sources within the FourStar field to determine the plate scale (0.1592$\pm$0.0001\arcsec/pix) and the orientation to the cardinal directions of a single frame, which is selected for use as an astrometric master.

{\em WISE : } W1639 was observed twice by WISE -- once as part of the {\em WISE All Sky} release and once again as part of the {\em WISE 3 Band Cryo} extended mission. We have extracted data from both these catalogs in a 600\arcsec\ region around W1639 using the IRSA gateway\footnote{\url{http://irsa.ipac.caltech.edu}}. The reported photometry and astrometry for W1639 in the {\em All Sky} release arises from 11 epochs spread over a period of less than 2 days centred on 2010 Mar 11 (UT), while that in the {\em WISE 3 Band Cryo} release arises from 18 epochs centred on 2010 Sep 7 (UT). Because these data spans are short, we treat each observation as a single epoch obtained at the mean MJD of the observations that make up each catalog entry (as obtained from the relevant Single Exposure Source Tables). The result is a set of co-ordinates on the system defined by the 2MASS Point Source Catalog \citep{skrutskie2006}. To enable a differential comparison with our FourStar images, these co-ordinates were tangent-projected about the {\em WISE All Sky} release position of W1639, scaled by the FourStar plate scale, and offset to place W1639 at the approximate location where it is actually observed in our FourStar images.

\subsection{Results}
The result is a set of observations in a FourStar pixel system, which we then subjected to standard differential astrometric processing \citep{tinney2003,tinney1996}. A set of 20 reference stars was chosen to surround W1639 in the field-of-view. We required all 20 stars to be present in both WISE epochs and all FourStar epochs. The FourStar observations transfer on to the FourStar master frame with typical root-mean-square (rms) residuals of 3-5\,mas. These astrometric precisions are for the reference stars, which in the FourStar data are typically much brighter than W1639 itself. Tests with background stars of similar magnitude to W1639 indicate that the FourStar astrometric precision for W1639 itself is $\pm$8-10\,mas, and these are the precisions that have been adopted in our astrometric solution. 

In contrast, the WISE data have residuals with an rms of 100\,mas. This is hardly surprising -- WISE is a small telescope with 6\arcsec\ resolution, so 0.1\arcsec\ astrometric precision for this reference frame is itself remarkable. W1639 is quite bright in the WISE data, and the precision achieved for the reference stars is representative of that expected for W1639 itself. The 2 year time baseline delivered by WISE means that it provides critical astrometric information, in spite of this lower precision. The Magellan data constrain on the parallax motion, while the WISE data constrain the proper motion. 

The resulting astrometric solution is shown in Fig. \ref{astrometry}. With only five independent epochs, and with two separate  instruments coming into play, we consider this astrometry to be preliminary. In particular, the formal uncertainties do not reflect potential systematic differences between the two instrument's co-ordinate systems. Nonetheless, we see that W1639 has a {\em very} large proper motion (3069$\pm$40\,mas\,yr$^{-1}$ at a PA=169.1$\pm$0.4$\deg$) which places it in the top 30 fastest moving stars in the Solar Neighbourhood. The formal parallax solution is 200$\pm$12\,mas.  With only five epochs in hand, this formal uncertainty will be an underestimate and the real uncertainty will be larger -- we recommend considering this preliminary parallax good to $\pm$20\,mas until a FourStar-only parallax can be produced within the next 12 months. Parallactic motion has, nonetheless, been detected with high significance, and we expect this solution to rapidly improve over the next 12 months.

\section{Discussion}

\begin{figure}
	\begin{center}
   \includegraphics[clip=true,width=7.0cm]{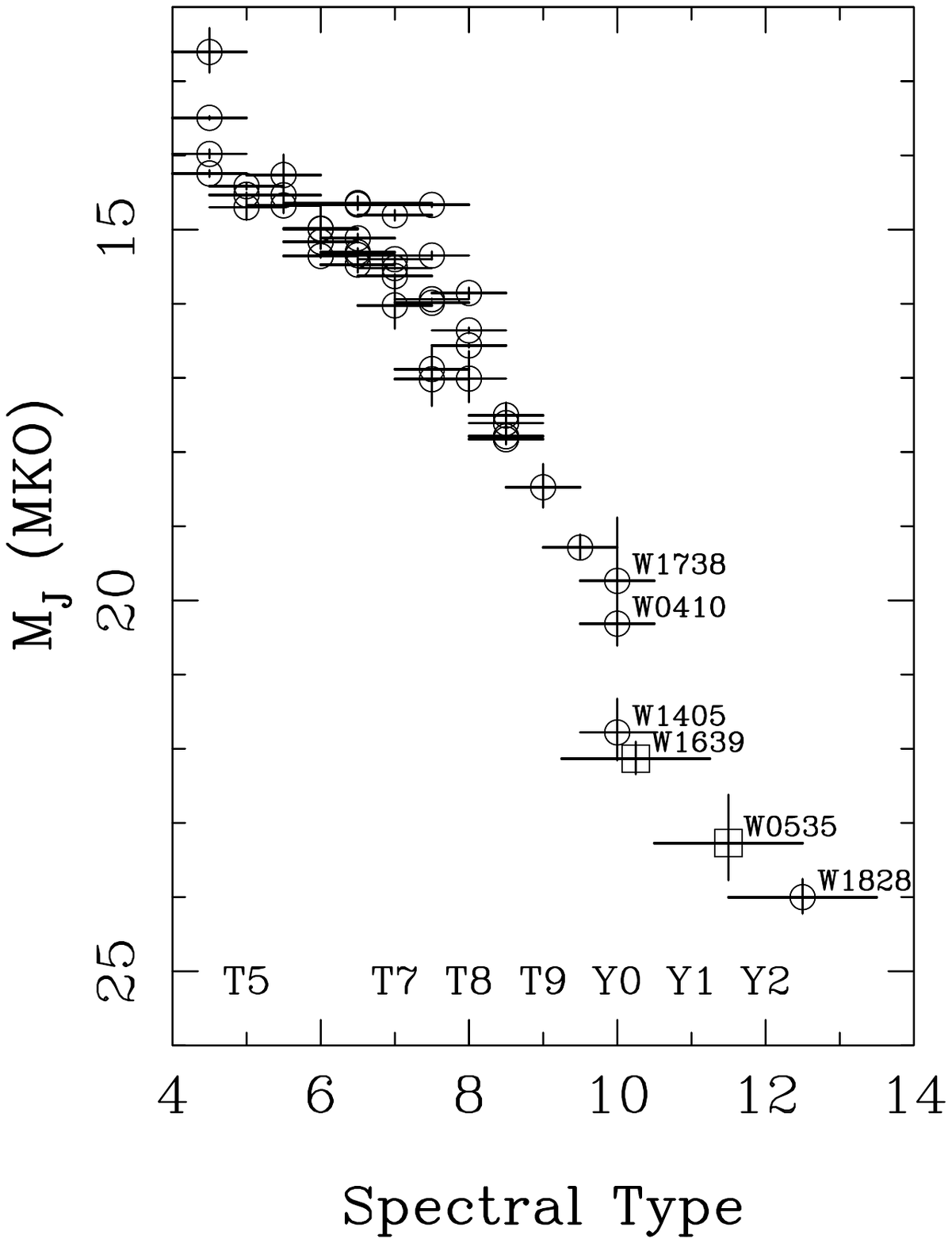}
      \caption{M$_{\mathrm J}$ absolute magnitude versus spectral type for late-T and Y dwarfs. Objects plotted are those with extant parallaxes (as summarised in Table 6 of \cite{kirkpatrick2012}) as well as the W1639 data presented in this paper. J photometry is on the Mauna Kea Observatories system -- objects from Kirkpatrick's table without MKO photometry available in the literature are not plotted. W1639 and WISE\,J053516.80$-$750024.9 have had J3 photometry converted to J as described in the text. Uncertainties on spectral types are plotted as $\pm$0.5 subtypes except (a) where Kirkpatrick et al. only assign an upper limit spectral type, in which case the sub-type is shifted by +0.5 and plotted with an uncertainty of $\pm$1 subtypes  (e.g. $>$Y1 is plotted as 11.5$\pm$1); and (b) the spectral type is noted as uncertain, in which case it is plotted as $\pm$1 subtype. W1639's Y0-Y0.5 type is plotted as 11.25$\pm$1.
      }
   \label{cmd}
	\end{center}
\end{figure}

W1639 has a measured distance of 5.0$\pm$0.5\,pc. This places it comfortably inside the 8pc Nearby Star Sample and makes it about the 55th closest stellar system to the Sun \citep{kirkpatrick2012}.  The measured distance to W1639 implies an absolute magnitude in the J passband (using the correction from J3 to J adopted earlier) of M$_{\mathrm J} = 22.14\pm0.22$. Figure \ref{cmd} places this result in context with other late-T and Y dwarfs with measured parallaxes  (spectral types are plotted as sub-type+10 for Y dwarfs, and sub-type+0 for T dwarfs. These data suggest that W1639 has an absolute magnitude in the J passband at the lower end of that observed for Y0 brown dwarfs (and just slightly fainter than that for the Y0 brown dwarf WISEPC\,J140518.40+553421.5), but significantly brighter than  the $>$Y1 brown dwarf WISE\,J053516.80$-$750024.9 \citep{kirkpatrick2012}. This is in line with the spectra presented earlier, which suggest a Y0-Y0.5 spectral type.

Combining the distance and proper motion, W1639 has a significant tangential velocity (V$_{tan}$) of 73$\pm$8\,\kms,  and is among the fastest moving normal (i.e. non-subdwarf or non-low-surface-gravity) cool brown dwarfs yet observed \citep{faherty2009,kirkpatrick2011,faherty2012}. \cite{faherty2012} derived the median V$_{tan}$ and $\sigma_{tan}$ values for the closely related T dwarf population to be 31\,\kms\ and 20\,\kms\ (respectively), suggesting that W1639 is kinematically deviant from brown dwarfs with ages typical of the field population (3-5 Gyr).   While individual velocities can not be used as an age indicator, general information can be obtained by comparing to the population of similar temperature objects. \cite{faherty2009} found that the members of the population of high-V$_{tan}$, low-temperature brown dwarfs show a correlation with being blue photometric outliers (and hence having  either low-metallicity, high surface gravity, thin photospheric clouds, or a combination thereof), and can be regarded as being kinematically older than the field population.  Using the measured distance, proper motion, and galactic coordinates ($l_{II}=321.2, b_{II}=-14.5$), we have estimated likely values for the full space motion for W1639 (assuming a radial velocity in a range similar to the V$_{tan}$ observed: $-$73, 0 and 73\,mms) and find a range of UVW velocities which consistently place W1639 outside  the ``Eggen box'' for young disk stars ($<$ 2 Gyr), indicating that this source is likely older than the field population \citep{eggen1989}.

While there are a limited number of Y dwarfs with distance measurements for comparison (Marsh et al. submitted, Beichman et al. submitted), \cite{kirkpatrick2012}report V$_{tan}$  values for a portion of newly discovered WISE brown dwarfs using photometric distance estimates for all sources.  Two Y dwarfs, WISE 0410+1502 (Y0) and WISE 1405+5534 (Y0pec) also have significant tangential velocities (V$_{tan} > 100$\,\kms) .   We note that this might be an indication that photometric distances are overestimated for these new sources.  More Y dwarfs with distance measurements are needed to investigate whether the kinematics of the new low temperature class of objects differs from that of the warmer brown dwarfs.

\section{Conclusion}

We have detected the near-infrared counterpart to the WISE mid-infrared source WISE\,J163940.83$-$684738.6. Spectroscopy indicates a spectral type of Y0-Y0.5, and this is consistent with the near-infrared-to-WISE J3--W2 colour of W1639 which suggests it to be of type $\approx$Y1. W1639 is a bright, nearby Y dwarf -- it is the brightest Y dwarf in the W2 passband in the WISE survey. It is likely to be a ``hot button'' target for future studies of the coolest brown dwarfs as its large proper motion moves it quickly away from the contaminating field star 2MASS\,J163940.83-684738.6 over the next 12 months.

These observations have shown, once again, that methane filters have significant utility in the follow-up of large surveys for cool brown dwarfs -- the use of imaging permits the sort of differential analysis that can identify Y dwarfs even when they lie quite close to bright background stars, while the unique methane signature provides an unambiguous means of identifying very cool photospheres. Moreover, we have shown that methane filters placed in the near-infrared J band (as installed in FourStar) can be very powerful for observations of the coolest brown dwarfs. At very cold temperatures, the J-band methane signature is significant (J2--J3 $\lesssim -1.0$), and allows observers to take advantage of the much fainter sky in the J-band -- especially compared to that seen by more traditional H-band methane filters \citep{tinney2005}. 

\section{Acknowledgments}

CGT gratefully acknowledges the support of ARC Australian Professorial Fellowship grant DP0774000. Australian access to the Magellan Telescopes was supported through the National Collaborative Research Infrastructure Strategy of the Australian Federal Government. Travel support for Magellan observing was provided by the Australian Astronomical Observatory.

This publication makes use of data products from the Wide-field Infrared Survey Explorer, which is a joint project of the University of California, Los Angeles, and the Jet Propulsion Laboratory/California Institute of Technology, funded by the National Aeronautics and Space Administration. This publication also makes use of data products from 2MASS, which is a joint project of the University of Massachusetts and the Infrared Processing and Analysis Center/California Institute of Technology, funded by the National Aeronautics and Space Administration and the National Science Foundation. This research has made extensive use of the NASA/IPAC Infrared Science Archive (IRSA), which is operated by the Jet Propulsion Laboratory, California Institute of Technology, under contract with the National Aeronautics and Space Administration. 

And finally, this research has benefitted from the M, L, and T dwarf compendium housed at DwarfArchives.org, whose server was funded by a NASA Small Research Grant, administered by the American Astronomical Society. We gratefully acknowledge the assistance of Phil Lucas and Ben Burningham who provided a digital copy of their UGPS\,0722 spectrum.\\

{\it Facilities:} \facility{Magellan:Baade (FourStar, FIRE)}, \facility{WISE}

\end{document}